\documentclass[12pt]{article}
\usepackage{epsfig,amssymb}
\renewcommand{\theequation}{\thesection.\arabic{equation}}

\begin{document}

\begin{flushright}
{\tt hep-th/0407229}
\end{flushright}

\vspace{5mm}

\begin{center}
{{{\Large \bf Noncommutative Tachyon Kinks\\[2mm]
as D$(p-1)$-branes from Unstable D$p$-brane}}\\[14mm]
{Rabin Banerjee}\\[3mm]
{\it S.~N.~Bose National Centre for Basic Sciences, Kolkata 700098, India}\\
{\tt rabin@bose.res.in}\\[7mm]
{Yoonbai Kim \, and \,  O-Kab Kwon}\\[3mm]
{\it BK21 Physics Research Division and Institute of
Basic Science,\\
Sungkyunkwan University, Suwon 440-746, Korea}\\
{\tt yoonbai@skku.edu~~okab@skku.edu} }
\end{center}
\vspace{10mm}

\begin{abstract}
We study noncommutative (NC) field theory of a real NC tachyon and
NC U(1) gauge field, describing the dynamics of an unstable
D$p$-brane. For every given set of diagonal component of open
string metric $G_{0}$, NC parameter $\theta_{0}$, and
interpolating electric field ${\hat E}$, we find all possible
static NC kinks as exact solutions, in spite of complicated NC
terms, which are classified by an array of NC kink-antikink and
topological NC kinks. By computing their tensions and charges,
those configurations are identified as an array of D0${\bar {\rm
D}}$0 and single stable D0 from the unstable D1, respectively.
When the interpolating electric field has critical value as
$G_{0}^{2}={\hat E}^{2}$, the obtained topological kink becomes a
BPS object with nonzero thickness and is identified as BPS D0 in
the fluid of fundamental strings. Particularly in the scaling
limit of infinite $\theta_{0}$ and vanishing $G_{0}$ and ${\hat
E}$, while keeping $G_{0}\theta_{0}={\hat E}\theta_{0}=1$,
finiteness of the tension of NC kink corresponds to tensionless
kink in ordinary effective field theory. An extension to stable
D$(p-1)$ from unstable D$p$ is straightforward for pure electric
cases with parallel NC parameter and interpolating two-form field.
\end{abstract}

\newpage

\setcounter{equation}{0}
\section{Introduction}

When the Neveu-Schwarz (NS) type background two-form field is
turned on, the corresponding string theory contains two mass
scales with a dimensionless string coupling $g_{{s}}$, i.e., they are
the string scale $\sqrt{\alpha'}$ and the magnitude of the
background field $1/\sqrt{|B|}$. Computation of propagator on a disc
in terms of boundary conformal field theory (BCFT) gives a relation between
closed and open string theory variables~\cite{Fradkin:1985qd}.
An intriguing aspect of Dirac-Born-Infeld (DBI)
limit of open string theories, containing many derivative terms,
is that the equivalence between the commutative spacetime theory
in terms of closed string variables and the noncommutative (NC)
spacetime analogue in terms of open string
variables~\cite{Seiberg:1999vs}. In the context of noncommutative field theory
(NCFT) with fixed
NC parameter $\theta$, the NC scale $\sqrt{|\theta|}$ replaces the
magnitude of the background field $1/\sqrt{|B|}$.
In case of the pure magnetic background $B_{ij}$, two scales,
$\theta^{ij}$ and $\alpha'$, in the NCFT can be
decoupled~\cite{Seiberg:1999vs}. On the other hand, with pure
electric background $B_{0i}$, an (NC) field theory limit of the
string theory is known to be not available in the limit of critical
electric field or equivalently $|\theta^{0i}|\gg \alpha'$ limit
due to possible problems like
unitarity~\cite{Seiberg:2000gc,Gopakumar:2000na}.
In the context of string theory, an appropriate scaling limit toward this
singular condition is known to lead to
NC open string theories (NCOS) and a theory of light open membranes
(OM)~\cite{Gopakumar:2000na,Gopakumar:2000ep}.

Another noteworthy example in NCFT is existence of static soliton
solutions, so-called GMS solitons~\cite{Gopakumar:2000zd}, identified
as D-branes and strings~\cite{Dasgupta:2000ft,Harvey:2000jt}.
Since almost all of the solitonic excitations
are naturally codimension-two objects
in (2$n$+1)-dimensions with spatial noncommutativity,
vortex-like configurations are obtained in various
NCFT's~\cite{Polychronakos:2000zm,Ghoshal:2004dd}, and they are used for
the description of decay to D0-branes from
D2${\bar {\rm D}}$2-system~\cite{Aganagic:2000mh}
through NC tachyon
condensation~\cite{Dasgupta:2000ft,Harvey:2000jt,Gopakumar:2000rw}.

It is well known that there also exist an unstable D$p$-brane
where $p$ is odd for type IIA string theory and even for IIB.
Its instability is represented by a real tachyonic degree, and it is
an intriguing issue to obtain stable codimension-one
D-brane from the unstable D$p$-brane~\cite{Sen:1998sm}.
In the context of effective field
theory (EFT) or NCFT, this question is translated as how to
obtain a stable static kink solution where the tension of D$(p-1)$-brane is
correctly computed~\cite{Sen:2003tm}. In EFT, various static
solitonic configurations of codimension-one including single kink and
array of kink-antikink are obtained, which are thin
or thick, with or without constant U(1) gauge field (equivalently NS-NS two
form field) for arbitrary
$p$~\cite{Sen:2003tm,Lambert:2003zr,Kim:2003in,Brax:2003rs,Kim:2003ma}.
Under a specific runaway tachyon potential, kinks identified by
codimension-one D-branes are given as
exact solutions~\cite{Lambert:2003zr,Kim:2003in} and, with critical electric
field, a thick topological kink is identified as a BPS
soliton~\cite{Kim:2003in}. Up to the present, such a rich
structure seems unlikely to be found in NCFT~\cite{Mandal:2000cx}.
In this paper, we will tackle this issue by combining the following
two wisdoms:
One is, in EFT, the action of real tachyon takes DBI
type~\cite{Garousi:2000tr} and the other is, in EFT and NCFT,
the NS-NS two-form field and U(1) gauge field are proven to share
the same DBI type actions connected by Seiberg-Witten (SW)
map~\cite{Seiberg:1999vs}.

We propose a new DBI type action of NC tachyon with coupling of NC
U(1) gauge field, describing dynamics of
an unstable D$p$-brane.
An important feature of this action is its equivalence (up to the first
non-trivial order in the NC parameter) with the corresponding DBI action of
tachyons in the commutative case. The equations of motion obtained from
our NC DBI action, for a flat unstable D1-brane with arbitrary diagonal
component of open string metric $G_{0}$ and interpolating electric field
$\hat{E}$, lead to static kink solutions. We abstracted all such solutions
as an array of NC kink-antikink and topological (NC) kinks. Furthermore,
when the interpolating electric field has the critical value $\hat{E}^2 = G_0^2$,
the topological kink reduces to a BPS object with non-vanishing thickness that is
identified as a BPS D0 in the fluid of fundamental strings.
Properties of the obtained kinks are consistent at on-shell level
with those in terms of
boundary string field theory~\cite{Gerasimov:2000zp},
EFT~\cite{Sen:2003tm,Lambert:2003zr,Kim:2003in,Brax:2003rs,Kim:2003ma}, and
BCFT~\cite{Sen:2003bc}. For the rolling tachyon solutions in the presence
of pure electric field, BCFT results are compared with those of NCFT
at the level of classical energy-momentum tensor~\cite{Mukhopadhyay:2002en}.

The rest of the paper is organized as follows. In section 2 we review
briefly DBI action of NC U(1) gauge field and then propose
an extension of this action such that it includes a NC tachyon field.
For slowly varying fields, equivalence between
ordinary effective action and NCFT action (in the presence of the tachyon)
is shown up to leading order
of the NC parameter; this is presented in Appendix A. In section 3,
the (1+1)-dimensional case is considered in detail.
All the static NC kink solutions are obtained as exact solutions.
They are identified as codimension-one branes by computing their
tensions and charges.
We conclude in section 4 with a brief discussion.

\setcounter{equation}{0}
\section{DBI Gauge Field and NC Tachyon}

In this section, we discuss our new proposal for the DBI type action of NC tachyon
coupled to NC U(1) gauge field. First, in subsection \ref{21}, we review
briefly both DBI action of U(1) gauge field and NS-NS
two-form field in the background of closed string metric, and that
of NC U(1) gauge field in the background of open string metric.
Various relations due to equivalence between those two actions are
also presented together with interpolating two-form field. Then,
in subsection \ref{22}, we propose a DBI type NC action of a real NC
tachyon coupled to NC U(1) gauge field based on the DBI type
action of ordinary effective field theory.
Equivalence of the two tachyon actions is shown up to the
leading order of NC parameter, for slowly varying fields.

\subsection{Review on DBI action of NC U(1) gauge field}\label{21}

DBI action describes slowly varying gauge field $A_{\mu}(x)$ on a
single D$p$-brane;
\begin{equation}\label{dbi}
S_{{\rm DBI}}=-\frac{1}{g_{{\rm s}}(2\pi)^{\frac{p-1}{2}}} \int
d^{p+1}x \sqrt{-{\rm det} (g_{\mu\nu}+B_{\mu\nu}+F_{\mu\nu})}\, ,
\end{equation}
where string coupling $g_{{\rm s}}$, metric $g_{\mu\nu}$, and a
constant NS-NS two-form field $B_{\mu\nu}$ are the closed string
variables read on the D$p$-brane. $F_{\mu\nu}$ denotes field
strength tensor of the U(1) gauge field, whose value is a
constant everywhere or vanishes at asymptotic region. In addition
to equation of the gauge field, it satisfies Bianchi identity
\begin{equation}\label{bia}
\partial_{\mu}F_{\nu\rho}+\partial_{\nu}F_{\rho\mu}+\partial_{\rho}
F_{\mu\nu}=0.
\end{equation}
Note that tension of the D$p$-brane ${\cal T}_{p}$ is inversely
proportional to the closed string coupling
\begin{equation}\label{ten}
{\cal T}_{p}=\frac{1}{g_{{\rm s}}(2\pi)^{\frac{p-1}{2}}}.
\end{equation}

A BCFT calculation of propagator on a disc, which corresponds to a
point splitting regularization of string theory, provides open
string metric $G^{\mu\nu}$ and NC parameter $\theta^{\mu\nu}$ in
terms of the closed string variables
as~\cite{Fradkin:1985qd} (see also
\cite{Seiberg:1999vs})
\begin{eqnarray}
G_{\mu\nu}&=&g_{\mu\nu}-(Bg^{-1}B)_{\mu\nu},
\label{omet}\\
\theta^{\mu\nu}&=&-\left(\frac{1}{g+B}B\frac{1}{g-B}\right)^{\mu\nu}.
\label{ncpa}
\end{eqnarray}
NC DBI type action, which is proven to be equivalent
to ordinary DBI action (\ref{dbi}) in the limit of slowly varying
fields~\cite{Seiberg:1999vs}, is given by
\begin{equation}\label{ncbi}
{\hat S}_{{\rm DBI}}= -\frac{1}{G_{{\rm
s}}(2\pi)^{\frac{p-1}{2}}}\int d^{p+1}x \sqrt{-{\rm det}
(G_{\mu\nu}+\Phi_{\mu\nu}+\hat{F}_{\mu\nu})}\, ,
\end{equation}
where NC field strength tensor ${\hat F}_{\mu\nu}$ is defined by
\begin{equation}\label{ncF}
\hat{F}_{\mu\nu}=\partial_{\mu}\hat{A}_{\nu}-\partial_{\nu}\hat{A}_{\mu}
-i\hat{A}_{\mu}\ast \hat{A}_{\nu} + i\hat{A}_{\nu}\ast
\hat{A}_{\mu},
\end{equation}
and $\Phi_{\mu\nu}$ is an interpolating two-form field depending
on $g_{\mu\nu}$, $B_{\mu\nu}$, and
$\theta_{\mu\nu}$~\cite{Seiberg:1999vs}.
Star product for NC fields is defined by
\begin{equation}
f(x)\ast g(x)\equiv e^{\frac{i}{2}\theta^{\mu\nu}
\frac{\partial}{\partial\xi^{\mu}}\frac{\partial}{\partial\zeta^{\nu}}}
f(x+\xi )g(x+\zeta)|_{\xi=\zeta=0}.
\end{equation}
NC Bianchi identity is the natural NC deformation of
the ordinary one (\ref{bia})~\cite{Banerjee:2003vc}
\begin{equation}\label{ncbia}
\hat{D}_{\mu} \hat{F}_{\nu\rho} +\hat{D}_{\nu} \hat{F}_{\rho\mu}
+\hat{D}_{\rho} \hat{F}_{\mu\nu}=0,
\end{equation}
where NC covariant derivative is
\begin{equation}\label{code}
\hat{D}_{\mu}= \partial_\mu - i [\hat{A}_\mu ,\;\;]_\ast \,
,\qquad [A,B]_\ast = A\ast B-B\ast A.
\end{equation}

When the field strength tensor $F_{\mu\nu}$ and the interpolating
field $\Phi_{\mu\nu}$ vanish, its NCFT analogue $\hat{F}_{\mu\nu}$
also vanishes so that the coupling of the open
string $G_{{\rm s}}$ is expressed by the closed string theory variables
\begin{equation}\label{cou}
G_{{\rm s}}=g_{{\rm s}}\sqrt{{\rm det}(1+g^{-1}B)}\, .
\end{equation}
Note that, from the coefficient of the quadratic term,
electromagnetic coupling $g_{{\rm EM}}$ is also identified as
\begin{equation}\label{gem}
\frac{1}{g_{{\rm EM}}^{2}}=\frac{1}{G_{{\rm
s}}(2\pi)^{\frac{p-1}{2}}}= \frac{1}{g_{{\rm s}}\sqrt{{\rm
det}(1+g^{-1}B)}(2\pi)^{\frac{p-1}{2}}}.
\end{equation}

For a given closed string metric with $\sqrt{-g}>0$, reality
condition of the actions Eqs.~(\ref{dbi}) and (\ref{ncbi}) is
presumed. Therefore, validity of the open string variables,
$G_{\mu\nu}$ (\ref{omet}), $\theta_{\mu\nu}$ (\ref{ncpa}), and
$G_{{\rm s}}$ (\ref{cou}), is justified by the critical line
characterized by $\sqrt{{\rm det}(1+g^{-1}B)}$ which is vanishing
DBI Lagrange density ${\cal L}_{{\rm DBI}}/\sqrt{-g}$ or
equivalently $G_{{\rm s}}/g_{{\rm s}}$ in Eq.~(\ref{cou}). Since
the DBI action (\ref{dbi}) is valid in weak string coupling limit
$(g_{{\rm s}}\rightarrow 0)$, it means $G_{{\rm s}}\rightarrow 0$
so does $g_{{\rm EM}}$ (\ref{gem}). Simultaneously, the open
string metric (\ref{omet}) vanishes, $\det G_{\mu\nu}\rightarrow
0$, and the NC parameter diverges, $\det\theta^{\mu\nu}\rightarrow
\infty$. It seems that the NCFT of pure $G_{\mu\nu}$ of our
interest becomes singular as the NS-NS field $B_{\mu\nu}$ approaches
critical value, and physically meaningless for the NS-NS field
$B_{\mu\nu}$ larger than the critical value.

\subsection{NC tachyon action}\label{22}

Let us begin this subsection by introducing an effective tachyon
action for the unstable D$p$-brane system in ordinary
spacetime~\cite{Garousi:2000tr}
\begin{equation}\label{fa}
S= -\frac{1}{g_{{{\rm s}}}(2\pi)^{\frac{p-1}{2}} } \int d^{p+1}x\;
V(T) \sqrt{-\det (g_{\mu\nu} + B_{\mu\nu}+F_{\mu\nu}+\partial_\mu
T\partial_\nu T)}\, .
\end{equation}
Since tachyon potential $V(T)$ measures variable tension of the unstable
D-brane, it can be any runaway potential connecting monotonically
\begin{equation}\label{vbd}
V(T=0)=1~~\mbox{and}~~ V(T=\pm\infty)=0.
\end{equation}
Physics of tachyon condensation is largely irrelevant to the
detailed form of the tachyon potential once it satisfies the
runaway property and the boundary values (\ref{vbd})~\cite{Sen:1999xm}.
For example,
both the basic runaway behavior of rolling tachyon
solutions~\cite{Sen:2002an}, existence of various tachyon kink
solutions~\cite{Kim:2003in,Kim:2003ma}, and their BPS nature with
zero thickness~\cite{Sen:2003tm} are attained irrespective of the
specific shape of the potential, which just reflects a detailed
decaying dynamics of the unstable D$p$-brane. So will be the NC
tachyon kinks in the context of NCFT, which will be shown in the
next section.

Here we also adopt a specific form of the tachyon potential
$V(T)$~\cite{Buchel:2002tj,Kim:2003he,Leblond:2003db} as
\begin{equation}\label{tpot}
V(T)=\frac{1}{\cosh (T/R)},
\end{equation}
where $R=\sqrt{2}$ for superstring theory and 2 for bosonic string
theory. This potential (\ref{tpot}) has some nice
features: (i) It is derived in open string theory by taking into
account the fluctuations around $\frac{1}{2}$S-brane configuration
with the higher derivatives neglected, i.e., $\partial^2 T =
\partial^3 T= \cdots = 0$~\cite{Kutasov:2003er}. (ii) Exact
solutions are obtained for rolling
tachyon~\cite{Kim:2003ma,Kim:2003he} and tachyon kink solutions on
unstable D$p$ without or with a coupling of U(1) gauge field for arbitrary
$p$~\cite{Lambert:2003zr,Kim:2003in,Kim:2003ma}.
(iii) Some of the obtained classical
solutions $T(x)$ in the EFT (\ref{fa}), e.g., rolling
tachyons and tachyon kinks, can be directly translated to BCFT
tachyon profiles $\tau(x)$ in open string
theory~\cite{Lambert:2003zr,Sen:2003bc} described by the following point
transformation obtained in Ref.~\cite{Kutasov:2003er},
\footnote{Intriguingly, such a transformation (for $R=1$) maps an Euclidean
sphere having radius $\tau$ with the corresponding hyperbolic
sphere having radius $T$ \cite{book}.}
\begin{equation}\label{poi}
\frac{\tau(x)}{R}=\sinh \left(\frac{T(x)}{R}\right).
\end{equation}
(iv) The period of the obtained array solutions of tachyon
kink-antikink~\cite{Lambert:2003zr,Kim:2003in,Kim:2003ma} or
tube-antitube~\cite{Kim:2003uc,Kim:2004xk} is independent of the
integration constant of the equation of motion only under this
potential (\ref{tpot})~\cite{Kim:2004xk}, which is a crucial
property in string theory if one wishes to identify the array
solution as a configuration on a circle or a sphere of a fixed
radius~\cite{Sen:2003bc,Sen:2003zf}. We will also demonstrate that
the properties (ii)--(iv) are indeed shared with the kinks in NCFT
(see the next section).

When the tachyon is considered as a real NC scalar field
in the context of NCFT, an action with quadratic kinetic term of it was
proposed~\cite{Dasgupta:2000ft} and has been used for studying
physics of unstable
D-branes~\cite{Dasgupta:2000ft,Harvey:2000jt,Aganagic:2000mh}. If we adopt such
action for NC tachyon, then the relation between it and the ordinary
tachyon action (\ref{fa}) may not be made clearly. In this paper,
we propose another NC tachyon action based on Eq.~(\ref{fa})
\begin{equation}\label{ncfa}
\hat{S}= -\frac{{\hat {\cal T}}_p}{2} \int d^{p+1}x\;
\left[\hat{V}(\hat{T}) \ast \sqrt{ -\hat X_\ast } +\sqrt{ -\hat X_\ast
}\ast \hat{V} (\hat{T}) \right].
\end{equation}
Each term in Eq.~(\ref{ncfa}) is obtained by replacing
 ordinary product with star product, {\it e.g.},
NC determinant in $(p+1)$-dimensional spacetime is defined as
\begin{eqnarray}\label{ncdet01}
\hat X_\ast =
{\det}_\ast \hat X_{\mu\nu} \equiv \frac{1}{(p+1)!}
\epsilon_{\mu_1\mu_2\cdots \mu_{p+1}} \epsilon_{\nu_1\nu_2\cdots
\nu_{p+1}} \hat X_{\mu_1\nu_1}\ast \hat X_{\mu_2\nu_2}\ast \cdots
\ast \hat X_{\mu_{p+1}\nu_{p+1}},
\end{eqnarray}
and full symmetrization is implied,
\begin{eqnarray}
\left[\hat A_1 \hat A_2 \cdots \hat A_n\right]_\ast &=&
\frac{1}{n!} \left(\hat A_1 \ast \hat A_2 \ast \cdots \ast \hat
A_n + \hat A_1 \ast \hat A_3  \ast \cdots \ast \hat A_n\right.
\nonumber\\
&& \left.\qquad + \cdots
(\mbox{all~possible~permutations})\right.\Big).
\label{stnot}
\end{eqnarray}
To be specific, we have
\begin{eqnarray}
\hat{V} (\hat{T}) &\equiv &  1 - \frac{1}{2} \frac{\hat T}{R}\ast
\frac{\hat T}{R} + \frac{5}{24}\frac{\hat T}{R}\ast
\frac{\hat T}{R}\ast \frac{\hat T}{R}\ast \frac{\hat T}{R} +\, \cdots
\nonumber \\
&=& 1 + \sum^{\infty}_{k=1} \frac{E_{2k}}{(2k)!} \left[
\left(\frac{\hat T}{R}\right)^{2k}\right]_\ast
= \left[ \frac{1}{\cosh ({\hat{T}}/R)}\right]_\ast,
\label{ncpot} \\
\hat X_\ast &=& {\det}_\ast \left[ G_{\mu\nu} + \hat F_{\mu\nu}
+ \frac{1}{2} \left(\hat D_\mu \hat T \ast \hat D_\nu \hat T +
\hat D_\nu \hat T \ast \hat D_\mu \hat T \right)\right],
\label{nX}
\end{eqnarray}
where $E_{2k}$ is the Euler number.
A quantity
\begin{eqnarray}
{\hat {\cal T}}_p \equiv\frac{1}{G_{{{\rm
s}}}(2\pi)^{\frac{p-1}{2}}}
=\frac{{\cal T}_{p}}{\sqrt{\det(1+g^{-1}B)}}
\end{eqnarray}
is introduced for convenient connection between the NC DBI action (\ref{ncfa})
and the ordinary DBI action (\ref{fa}) in the limit of vanishing
tachyon, $\hat T = 0$.

The aforementioned procedure of star products between all the fields and their
full symmetrization is not unique but seems likely to be a natural choice
in this stage.\footnote{This is a valid symmetrization since it is similar
to the familiar Weyl ordering prescription which is compatible with
the canonical definition of star product taken here.}
This kind of ambiguity is genuine even in usual NC scalar field
theory with equal to or more than $\phi^{6}$-potential
term~\cite{Aref'eva:2000hq},
and affects much on solitonic spectra, particularly on
codimension-two objects~\cite{Kiem:2001ny}.
However, note that the kink solutions, which will be dealt in the next section,
are supported irrespective of such detailed procedure for obtaining
the NC tachyon action once it takes a DBI type.
For slowing varying $\hat F_{\mu\nu}$ and $\hat D_\mu \hat T$, the
star products in $\hat X_\ast$ may be replaced by the ordinary products
according to the approach of Ref.~\cite{Seiberg:1999vs}.
Then the NC DBI action (\ref{ncfa}) is simplified as
\begin{equation}\label{ncfa-1}
\hat{S}= - {\hat {\cal T}}_p \int d^{p+1}x\;
\left[\hat{V}(\hat{T}) \sqrt{- \hat X} + {\cal O} (\partial \hat F,\,
\partial \hat D \hat T)\right],
\end{equation}
where all the products in the determinant are ordinary products
\begin{equation}
\hat X = \det (G_{\mu\nu} + \hat F_{\mu\nu}
+ \hat D_\mu \hat T \hat D_\nu \hat T).
\end{equation}
Now star products are contained only in the tachyon potential
$\hat V(\hat T)$. Since the DBI type effective action of tachyon can be
valid only for slowly varying tachyon and U(1) gauge field, it is enough
to prove up to leading order of the NC parameter,
the equivalence with the corresponding DBI action in the commutative case.
This would also establish consistency with \cite{Seiberg:1999vs}
where an analogous demonstration was carried out for the non-tachyonic theory.
Since it is lengthy, we give the detailed proof
in Appendix A.
The total derivative difference in Lagrange density
arises from the fact that both the
NC gauge field action and the NC tachyon action of our interest
can be derived in string theory at on-shell,
which are not sensitive to such total derivatives. The difference by
${\cal O}(\partial F)$ or ${\cal O}(\partial \hat D T)$ in the Lagrange density
is expected from the beginning
since such terms are neglected when the Lagrange densities (\ref{ncfa-1}) and
(\ref{fa}) are
derived for both NC and ordinary field theory cases.

Equations of motion for the NC tachyon and U(1) gauge field are given by
\begin{eqnarray}
&& \hat D_\mu \left(
\frac{\hat V (\hat T)}{\sqrt{- \hat X}}\,
\hat C^{\mu\nu}_{{\rm S}} \hat D_\nu\hat T\right)
- \left[ \frac{\sinh (\hat T/R) }{R
\cosh^2 (\hat T/R)}\right]_\ast \sqrt{ - \hat X} =0,
\label{ncte} \\
&& \hat D_\mu\left( \frac{\hat V(\hat T)}{\sqrt{-\hat X}} \,
\hat C^{\mu\nu}_{{\rm A}}\right) + i \left[ \hat T\, ,
\, \frac{\hat V(\hat T)}{
\sqrt{-\hat X}} \, \hat C^{\mu\nu}_{{\rm S}}\hat D_\nu \hat T \right]_\ast = 0,
\label{ncge}
\end{eqnarray}
where $\hat C^{\mu\nu}_{{\rm S}}$ and $\hat C^{\mu\nu}_{{\rm A}}$ are symmetric
and antisymmetric parts of the cofactor $\hat C^{\mu\nu}$ of the matrix
$(\hat{X})_{\mu\nu}$, computed as
\begin{equation}\label{dsX}
\delta_{\hat T} \sqrt{- \hat X} = - \frac{\delta_{\hat T} \hat X}{
2 \sqrt{- \hat X}} = - \frac{ \hat C^{\mu\nu}}{2 \sqrt{ - \hat X}}
\, \delta_{\hat T} \hat X_{\mu\nu}.
\end{equation}
The detailed procedure deriving the equations (\ref{ncte})--(\ref{ncge})
from the NC action (\ref{ncfa}) is given in Appendix B.

NC version of the energy-momentum tensor is read by a systematic
way~\cite{Banerjee:2003vc} as follows
\begin{eqnarray}\label{emte}
\hat{T}^{\mu\nu}&\equiv &  \frac{2}{\sqrt{-G}}
\frac{\delta\hat{S}}{\delta G_{\mu\nu}} = \frac{\hat {\cal T}_p \hat V
\hat C^{\mu\nu}_{{\rm S}}}{ \sqrt{-G} \sqrt{-\hat X}}
\end{eqnarray}
which is conserved covariantly
\begin{eqnarray}\label{nccem}
\hat{D}_{\mu} \hat{T}^{\mu\nu}=0.
\end{eqnarray}

\setcounter{equation}{0}
\section{Codimension-one Branes from Unstable D1-brane}

We consider a flat unstable D1-brane in NCFT having NC parameter $\theta_{0}$ and
diagonal component of open string metric $G_{0}$ with interpolating
electric field ${\hat E}$. In this section, we find all the static
solitonic objects of codimension-one, interpreted as an array of
D0$\bar{{\rm D}}$0 or D0-brane.

Suppose the closed string metric has only diagonal components and
 the antisymmetric tensor field has electric components as
\begin{eqnarray}
\left(g_{\mu\nu}\right) &=& \left(\begin{array}{ccc}
-g_0 & 0 \\0& g_0 \end{array}\right), \\
\left(B_{\mu\nu}\right) &=& \left(\begin{array}{ccc} 0 & E_0 \\
-E_0 & 0\end{array}\right),
\end{eqnarray}
where $g_0$ and $E_0$ are constants, and $g_{0}=1$ does not lose
generality in our discussion. By comparing the actions for
$p=1$, Eq.~(\ref{dbi}) with $F_{\mu\nu}=0$ and Eq.~(\ref{ncbi})
with ${\hat F}_{\mu\nu}=0$,
we read the open string coupling $G_{{\rm s}}$ (\ref{cou}) (or equivalently
${\hat {\cal T}}_{1}$ in NCFT) and metric
\begin{eqnarray}\label{opG}
\left(G_{\mu\nu}\right) &=& \left(\begin{array}{cc} -G_0 & 0 \\ 0
& G_0\end{array}\right).
\end{eqnarray}
Specifically, for nonnegative $g_{0}$ and $g_0^2-E_0^2$, we naturally have
\begin{eqnarray}\label{opc+}
G_0 = \frac{g_0^2- E_0^2}{g_0}\ge 0,\quad \frac{G_{{\rm
s}}}{\sqrt{G_{0}}}= \frac{g_{{\rm s}}}{\sqrt{g_{0}}}
~~(\sqrt{G_{0}}{\hat {\cal T}}_{1}=\sqrt{g_{0}}{\cal T}_{1}),
\end{eqnarray}
while the NC parameter has the form
\begin{eqnarray}
\left(\theta^{\mu\nu}\right) =\left(\begin{array}{cc} 0 & \theta_0
\\ -\theta_0 & 0\end{array}\right), \qquad \theta_0 =
\frac{E_0}{g_0^2-E_0^2}.
\label{opt}
\end{eqnarray}

We are interested in D0-branes from the unstable D1-brane, given as
static solitonic configurations. Therefore, NC fields are assumed
to depend only on $x$-coordinate, i.e., $\hat{T}= \hat{T}(x)$ and
$\hat{F}_{\mu\nu} = \hat{F}_{\mu\nu}(x)$. If we choose a Weyl
gauge $\hat A_0 =0$ for convenience, we have simplified
expressions as follows
\begin{eqnarray}
&&\hat F_{01} \equiv \hat E = \partial_0\hat A_1 - \partial_1 \hat
A_0 -i[\hat A_0, \hat A_1] = \partial_0 \hat A_1,
\nonumber \\
&&\hat T \ast \hat T = \hat T^2,\quad \hat D_\mu \hat
T = \delta_{\mu 1} \hat T ' (1 + \hat E \theta_0), \quad
\hat D_\mu \hat T \ast \hat D_\nu \hat T = \delta_{\mu 1}
\delta_{\nu 1} \hat T'^2 (1 + \hat E \theta_0)^2,
\label{tat}
\end{eqnarray}
where $ \hat T' = d \hat T/d x$. Remarkably, under this gauge,
every star product in the NC DBI action (\ref{ncfa}) and the NC equations
of motion (\ref{ncte})--(\ref{ncge}) is replaced
by ordinary product.\footnote{It appears therefore that the Weyl gauge
``abelianizes'' the NC theory just as the axial gauge (say $A_1\approx0$)
``abelianizes'' the usual Yang--Mills theory, allowing for a solution of
the Gauss constraint.} In this sense, the codimension-one objects of our interest
are insensitive to the way how to attach the star product, as previously mentioned.
This property (\ref{tat}) allows application of the same
point transformation (\ref{poi}) to the NC tachyon field ${\hat T}$.
Performing this transformation
(\ref{poi}) to ${\hat T}$, we have
${\hat \tau}\left(=R\sinh ({\hat T}/R)\right)$ and an NC action
\begin{equation}\label{ncfa01}
\hat S = -\hat {\cal T}_1  \int d^{2} x\,\hat V(\hat\tau) \sqrt{-
\hat X},
\end{equation}
where the tachyon potential (\ref{ncpot}) and the NC determinant (\ref{nX})
become
\begin{eqnarray}
\hat V(\hat\tau) &=&\frac{1}{\sqrt{1 + \hat \tau^2/R^2}},
\label{V01} \\
-{\hat{X}}_{\ast}=
-\hat X &=& G_0^2 - \hat E^2  + G_0 (1 + \hat E\theta_0)^2 \hat
V^2 \hat \tau'^2 . \label{X01}
\end{eqnarray}

Conjugate momentum of the gauge field ${\hat \Pi}$ defined by
\begin{equation}\label{hpi}
{\hat \Pi}\equiv \frac{\delta {\hat S}}{\delta {\hat F}_{01}}
=\frac{{\hat {\cal T}}_{1}{\hat V}}{\sqrt{-{\hat X}}} {\hat E}
\end{equation}
is time independent due to spatial component of the gauge field
equation (\ref{ncge}), i.e., $\partial_{0}{\hat \Pi}=0$ under the
Weyl gauge. Then the equations of motion for the tachyon (\ref{ncte})
and Gauss' law constraint, time component of the gauge field
equation (\ref{ncge}), are
\begin{eqnarray}
\lefteqn{(1 + \hat E \theta_0)\left[ \frac{\hat {\cal T}_1 \hat
V}{ \sqrt{- \hat X}} \hat V^2 G_0 (1 + \hat E \theta_0)  \hat
\tau'\right]^{'}
}\nonumber\\
&&+ \frac{\hat {\cal T}_1 \hat V }{\sqrt{-\hat X}} \frac{{\hat
V}^2 \hat \tau}{R^2} \left[ -\hat X + G_0 \hat V^2 (1 + \hat
E\theta_0)^2\hat \tau'^2\right] = 0,
\label{ncte01}\\
&&(1 + \hat E \theta_0)({\hat \Pi})^{'}= (1 + \hat E \theta_0)
\left( \frac{\hat {\cal T}_1 \hat V}{\sqrt{- \hat X}}\hat
E\right)^{'}=0. \label{ncge01}
\end{eqnarray}
Instead of solving the complicated tachyon equation
(\ref{ncte01}), it is convenient to consider $x$-component of the
conservation of energy-momentum (\ref{nccem})
\begin{equation}\label{nccem01}
\hat D_1 \hat T^{11} = (1 + \hat E \theta_0) \left(\frac{-\hat
{\cal T}_1 \hat V}{\sqrt{- \hat X}}\right)^{'} =0.
\end{equation}

When $1 + \hat E \theta_0$ vanishes, the derivative terms of the
tachyon field disappear as shown in Eq.~(\ref{tat}) and thereby
the equations of motion (\ref{ncte01})--(\ref{nccem01}) become
trivial. So no nontrivial solitonic object is obtained except for
trivial vacuum solutions, ${\hat \tau}=0$ or ${\hat
\tau}=\pm\infty$.

When $1 + \hat E \theta_0 \ne 0$, Eq.~(\ref{nccem01}) dictates
constancy of $\hat T^{11}$ as
\begin{equation} \label{nccem02}
- \hat T^{11} = \frac{\hat {\cal T}_1 \hat V}{\sqrt{-\hat X}}.
\end{equation}
Then the gauge equation (\ref{ncge01}) allows only constant electric
field, $\hat E = \mbox{constant}$, and simultaneously it means
that constancy of conjugate momentum to the gauge field, ${\hat
\Pi}=-\hat T^{11}{\hat E}$ due to Eq.~(\ref{hpi}).
Momentum density $\hat T^{01}$ vanishes, and then energy density
is the only component of energy-momentum tensor with nontrivial
profile
\begin{equation}\label{t00}
-{\hat T}^{0}_{\; 0}={\hat \Pi}\frac{{\hat E}}{G_{0}}+ \frac{{\hat
E}/G_{0}}{{\hat \Pi}}( {\hat {\cal T}}_{1}{\hat V} )^{2}.
\end{equation}

Obviously these solutions satisfy the tachyon equation
(\ref{ncte01}). Finally, Eq.~(\ref{nccem02}) is rewritten as
\begin{equation}\label{fde}
{\cal E}_1 = \frac{1}{2} \hat\tau'^2  + U_1(\hat\tau),
\end{equation}
where
\begin{eqnarray}
{\cal E}_1 &=& \frac{1}{2}\left[ \frac{{\hat {\cal
T}}_1^2}{G_{0}(-{\hat T}^{11}(1 + \hat E\theta_0))^2} -\frac{G_0^2
- \hat E^2}{G_{0}(1 + \hat E\theta_0)^2} \right],
\label{enE} \\
U_1(\hat\tau) &=&  \frac{1}{2} {\hat \omega}^2  R^{2}
\left(\frac{1}{{\hat V}^{2}} -1\right), \qquad \hat \omega =
\frac{1}{R}\sqrt{\frac{G_0^2 - \hat E^2}{G_0(1+\hat
E\theta_0)^2}}\, .
\label{potU}
\end{eqnarray}
For a given geometry with a fixed $g_{0}\ne 0$ (or equivalently a
fixed $G_{0}$) and the compactification scale $R$, the system of
our interest seems to be classified by three parameters, i.e.,
they are negative pressure $-{\hat T}^{11}$, NC parameter
$\theta_{0}$, and NC electric field $\hat E$. On the other hand,
the solution space of static codimension-one objects is classified
by two parameters ${\cal E}_1$ and $\hat \omega$.
Specifically, the following two combinations are read from
Eqs.~(\ref{enE})--(\ref{potU})
\begin{equation}
\frac{{\hat {\cal T}}_1^2}{G_{0}[-{\hat T}^{11}(1 + \hat
E\theta_0)]^2}, \qquad \frac{G_0^2 - \hat E^2}{G_0(1+\hat
E\theta_0)^2}.
\end{equation}
In fact, value of the NC electric field $\hat E$ is constant
version of two-form auxiliary field $\Phi_{\mu\nu}$ in Eq.~(\ref{ncbi}).
It
interpolates between the limit of ordinary EFT with a
constant electric field $E$ for $(\theta_{0}=0,E={\hat E})$ and
that of NCFT with $E/g_{0}=G_{0}\theta_{0}$ for $(\theta_{0}\ne
0,{\hat E} =0)$. Since the analogue of $\hat \omega$ in ordinary
effective theory is
$\omega=\sqrt{(g_{0}^{2}-E^{2})/g_{0}}/R$~\cite{Kim:2003in},
comparison with Eq.~(\ref{potU}) provides a relation to have
identical configurations at intermediate values
\begin{equation}
1-\left(\frac{E}{g_{0}}\right)^{2}
=\frac{\left[1-\left(G_{0}\theta_{0}\right)^{2}\right]^{2}
-\left(\frac{{\hat E}}{g_{0}}\right)^{2}}{
\left[1-\left(G_{0}\theta_{0}\right)^{2}\right]
\left[1+\frac{G_{0}\theta_{0}\frac{{\hat E}}{g_{0}}}{
1-\left(G_{0}\theta_{0}\right)^{2}}\right]^{2}}.
\end{equation}

We examine the equation (\ref{fde}) by dividing the cases into
three, i.e., ${\hat \omega}^{2}>0$,  ${\hat \omega}^{2}=0$, and
${\hat \omega}^{2}<0$. As a boundary condition of ${\hat
\tau}(x)$, we use $\hat\tau(0) = 0$ without losing generality.

\subsection{Array of kink and antikink for ${\hat \omega}^{2}>0$}
When ${\hat \omega}^{2}>0$, nontrivial solution can exist for
positive ${\cal E}_{1}$ which leads to ${\hat {\cal T}}_1^2/(G_0^2
- \hat E^2)(-{\hat T}^{11})^{2}> 1$, and it is an oscillating
configuration for any runaway tachyon potential $\hat V$ with
${\hat V} ({\hat \tau}=0)=1$ and ${\hat V} ({\hat \tau}=\pm\infty)
=0$. Under the specific potential (\ref{ncpot}), we obtain an
exact solution
\begin{equation}\label{oscs}
\frac{\hat\tau (x)}{R} = \pm \sqrt{\frac{{\hat {\cal T}}_1^2}{
(G_0^2 - \hat E^2)(-{\hat T}^{11})^{2}} - 1}\, \sin (\hat \omega
x).
\end{equation}
The NC tachyon field oscillates sinusoidally between maximum and
minimum values, $\pm R\sqrt{\frac{{\hat {\cal T}}_1^2}{ (G_0^2 -
\hat E^2)(-{\hat T}^{11})^{2}} - 1}$ with period $2\pi/{\hat
\omega}=2\pi R\sqrt{G_0 (1 + \hat E\theta_0)^2/(G_0^2 - \hat
E^2)}\,$. Then this solution is interpreted as the array of kink
and antikink in the presence of NC electric field $\hat E$ transverse to
the kink (or antikink). Note that the period does not depend on the
an integration constant $-{\hat T}^{11}$. This property is unique under the
specific form of tachyon potential (\ref{V01}). A proof is simple
as given in the following. Once we identify the system described
by Eq.~(\ref{fde}) as that of a hypothetical particle with unit
mass, of which position is ${\hat \tau}$ and time $x$, then
possible motion is nothing but that of a one-dimensional simple
harmonic oscillator and its period $2\pi/{\hat \omega}$ is
independent of mechanical energy ${\cal E}_1$. Such
${\cal E}_1$-independency holds uniquely for the simple harmonic oscillator as
proved in Ref.~\cite{LLt}.

This phenomenon can also be seen in the
NC action (\ref{ncfa01}) through a rescaling of spatial coordinate
$x\rightarrow \chi =\sqrt{(G_0^2 - \hat E^2)/G_0 (1 + \hat
E\theta_0)^2}\, x$
\begin{eqnarray}
{\hat S}&=&-\sqrt{G_0}{\hat {\cal T}}_{1}(1 + \hat E\theta_0) \int
dt dx \, {\hat V}\sqrt{\frac{G_0^2 - \hat E^2}{G_0 (1 + \hat
E\theta_0)^2} +{\hat V}^{2}{\hat \tau}'^{2}}
\label{act2}\\
&=&-\sqrt{G_0}{\hat {\cal T}}_{1} (1 + \hat E\theta_0) \int dt
d\chi \, {\hat V}\sqrt{1+{\hat V}^{2}\left(\frac{d{\hat
\tau}}{d\chi} \right)^{2}}\, . \label{reac}
\end{eqnarray}
Formal resemblance between Eq.~(\ref{reac}) and the rescaled
action in ordinary effective action with constant electric field
is clear under the point transformation (\ref{poi}) and the
relation (\ref{tat})~\cite{Kim:2003in}. In the pure NCFT limit
with vanishing ${\hat E}$, an exact identification is made by the
relation between ${\cal T}_{1}$ and ${\hat {\cal T}}_{1}$ (\ref{opc+}).

Substituting the equation of motion (\ref{fde}) and kink solution
(\ref{oscs}) into the NC action (\ref{act2}) for half period, we
obtain the formula for tension of a unit kink (or unit antikink)
\begin{eqnarray}
\frac{{\hat S}}{-\int dt\sqrt{G_{0}}}&=& \frac{{\hat {\cal
T}}_{1}^{2} }{-{\hat T}^{11}\sqrt{G_{0}}} \int_{-\frac{\pi}{2{\hat
\omega}}}^{\frac{\pi}{2{\hat \omega}}} dx\,
{\hat V}^{2}
\label{g0}\\
&=&\pi R {\hat {\cal T}}_{1} (1 + \hat E\theta_0)
\label{art0}\\
&\equiv& {\hat {\cal T}}_{0} (1 + \hat E\theta_0), \label{arte}
\end{eqnarray}
where a relation from the open string metric (\ref{opG}),
\begin{equation}
\int dt dx \sqrt{-\det (G_{\mu\nu})}=\int dt dx G_{0}
=\int dt\sqrt{G_{0}} \times \int dx \sqrt{G_{0}},
\end{equation}
was used in the left-hand side of Eq.~(\ref{g0}) and will be used
in the formulas of Hamiltonian and tension.
The tension is corrected by a factor $(1+ \hat E\theta_0)$ when
the electric field $\hat E$ exists in NC spacetime, however it
reduces to the value ${\hat {\cal T}}_{0}$ in the limit of either vanishing
electric field $({\hat E}\rightarrow 0)$ or pure commutative
spacetime $(\theta_{0}\rightarrow 0)$. The obtained decent
relation in NCFT (\ref{art0})--(\ref{arte})
is identical to ${\cal T}_{0}=\pi R {\cal T}_{1}$ in ordinary EFT
irrespective of the value of ${\hat E}$ and $\theta_{0}$. The
Hamiltonian for a single kink (or antikink) in the array is
\begin{eqnarray}
H&\equiv &\int^{\frac{\pi}{2{\hat \omega}}}_{-\frac{\pi}{2{\hat
\omega}}}
dx\sqrt{G_{0}}\, (-{\hat T}^{0}_{\;0})\\
&=& \frac{\pi}{{\hat \omega}}\sqrt{G_{0}} \, {\hat \Pi}\frac{{\hat
E}}{G_{0}} +\pi R{\hat {\cal T}}_{1}(1 + \hat E\theta_0).
\label{arha}
\end{eqnarray}
Comparing it to the tension from the action (\ref{arte}), the second term in
Eq.~(\ref{arha}) is the tension of a single D0-brane
(or $\bar {\rm D}_0$-brane)
and the first term comes from F1 string fluid of which the signal appears
through nonvanishing NC electric field ${\hat E}$.

Substituting the array solution (\ref{oscs}) into the energy
density (\ref{t00}), we have
\begin{equation}\label{sine}
-{\hat T}^{0}_{\;0}={\hat \Pi}\frac{{\hat E}}{G_{0}} +\frac{{\hat
E}/G_{0}}{{\hat \Pi}} \frac{{\hat {\cal T}}_{1}^{2}}{
1+\left[\frac{{\hat {\cal T}}_{1}^{2}}{{\hat \Pi}^{2}\left[\left(
\frac{G_{0}}{{\hat E}}\right)^{2}-1\right]}-1\right]
\sin^{2}\left[\sqrt{\frac{G_{0}^{2}-{\hat E}^{2}}{G_{0}(1+{\hat E}
\theta_{0})^{2}}}\,\frac{x}{R}\right] }
\end{equation}
composed of a constant density from the first string fluid term
and an oscillating contribution from the second array term, having values
between $\hat \Pi \left[1-({\hat E}/G_{0})^{2}\right] /({\hat
E}/G_{0})$ and ${\hat {\cal T}}_{1}^{2}({\hat E}/G_{0})/{\hat
\Pi}$.

From now on, let us take various limits. (i) Taking the NC
parameter to be zero $(\theta_{0}\rightarrow 0)$ keeping
${\hat E}$ and ${\hat \Pi}$ fixed, we obtain the limit of EFT with
$G_{0}=g_{0}$ smoothly~\cite{Kim:2003in,Kim:2003ma}. (ii) Turning
off the NC electric field $({\hat E}\rightarrow 0)$ keeping
$G_{0}$, $\theta_{0}$, and $-{\hat T}^{11}$ fixed, we easily
confirm disappearance of F1 contribution in pure NCFT limit
\begin{equation}
\left. -{\hat T}^{0}_{\;0}\right|_{{\hat E}\rightarrow 0} ={\hat
{\cal T}}_{1}\frac{\left(\frac{{\hat {\cal T}}_{1}}{-{\hat T}^{11}
G_{0}}\right)}{1+\left[ \left(\frac{{\hat {\cal T}}_{1}}{-{\hat
T}^{11} G_{0}}\right)^{2}-1
\right]\sin^{2}\left(\sqrt{G_{0}}\frac{x}{R}\right)}.
\end{equation}
It is consistent with vanishing F1 charge density in this limit,
i.e., $\lim_{\hat E \to 0} \hat \Pi = \lim_{\hat E \to 0} (-{\hat
T}^{11}) \hat E$ $\stackrel{{\rm finite}~ (- {\hat T}^{11})}{=}
0$. (iii) The limit of zero thickness is achieved by taking ${\hat
\Pi}\rightarrow 0$ with fixed $G_{0}$ and ${\hat E}$, where the
energy density (\ref{sine}) from the NC action (\ref{ncfa})
is given by a sum of $\delta$-functions
\begin{equation}\label{del}
\left. -{\hat T}^{0}_{\;0}\right|_{{\hat \Pi}\rightarrow 0} ={\hat
{\cal T}}_{0}(1+{\hat E}\theta_{0})\sum_{n=-\infty}^{\infty}
\frac{1}{\sqrt{G_{0}}}\delta\left(x-\frac{n\pi}{{\hat
\omega}}\right).
\end{equation}
Since ${\hat \Pi}\rightarrow 0$ limit with fixed ${\hat E}$ is
equivalent to vanishing pressure limit $-{\hat T}^{11}\rightarrow
0$ and each $\delta$-function in Eq.~(\ref{del}) stands for the
energy density of each kink or antikink, this zero thickness limit
corresponds to BPS limit in NCFT, which was the case in ordinary
EFT~\cite{Sen:2003tm,Kim:2003in}. From Eq.~(\ref{oscs}), each kink
becomes a topological kink connecting two true vacua at ${\hat
\tau}=\pm\infty$ though the configuration is singular. Note that
we have used ${\hat V}(0)=1$ and ${\hat V}(\pm\infty)=0$ to get
the BPS limit, which implies that it is saturated for any runaway
potential~\cite{Sen:2003tm}.

\subsection{Thick topological BPS kink for ${\hat \omega}^{2}=0$}

When ${\hat \omega}^{2}=0$, a drastic change is made due to
disappearance of the potential term $U_{1}=0$ in Eq.~(\ref{fde}),
and then we find a linear solution
\begin{equation}\label{lis1}
{\hat \tau}(x)= \pm\frac{{\hat {\cal T}}_1}{\sqrt{G_{0}} (-{\hat
T}^{11}) (1+{\hat E}\theta_{0})}~x,
\end{equation}
where ${\hat \tau}(0)=0$. Therefore, the obtained solution can
also be understood as the infinite period limit of the periodic
solution (\ref{oscs}) and then it describes single topological
kink (or antikink) connecting two vacua at ${\hat \tau}=\infty$
and ${\hat \tau}=-\infty$ smoothly. Note that, for finite $E_{0}$
and ${\hat E}$, formula of ${\hat \omega}$ (\ref{potU}) leads to
\begin{equation}\label{lim}
G_0^2 = \hat E^2
\end{equation}
so that the solution (\ref{lis1}) is rewritten by
\begin{equation}\label{lis2}
{\hat \tau}(x)= \pm\frac{{\hat {\cal T}}_1 \sqrt{{\hat E}}}{{\hat
\Pi} (1+{\hat E}\theta_{0})}~x.
\end{equation}

Let us read the tension of this kink or antikink from the action
(\ref{act2}) as follows:
\begin{eqnarray}
\frac{{\hat S}}{-\int dt\sqrt{G_{0}}}&=& {\hat {\cal T}}_{1}(1 +
\hat E\theta_0) \int_{-\infty}^{\infty} dx\, {\hat V}({\hat
\tau})\sqrt{{\hat \omega}^{2}R^{2}+
{\hat V}^{2}{\hat \tau}'^{2}} \\
&\stackrel{{\hat \omega}^{2}=0}{=}& {\hat {\cal T}}_{1}(1 + \hat
E\theta_0)
\int_{-\infty}^{\infty} d{\hat \tau} \, {\hat V}({\hat \tau})^{2} \\
&\stackrel{{\hat \tau}/R=\sinh ({\hat T}/R)}{=}& {\hat {\cal
T}}_{1}(1 + \hat E\theta_0) \int_{-\infty}^{\infty} d{\hat T}\,
V({\hat T})
\label{toch}\\
&=& {\hat {\cal T}}_{0}(1 + \hat E\theta_0), \label{tote}
\end{eqnarray}
where we used Eq.~(\ref{V01}) for ${\hat V}({\hat \tau})$ in the
second line and Eq.~(\ref{tpot}) for $V({\hat T})$ in the third
line.
The expression (\ref{toch}) coincides with the exact
integral formula of the tension for both the singular BPS
kink~\cite{Sen:2003tm} and the thick BPS kink with critical
electric field~\cite{Kim:2003in,Kim:2003ma,Kwon:2003qn}.

For the topological kink, the pressure ${\hat T}^{11}$ is provided
by a constant background of the F1, $- {\hat T}^{11}={\hat
\Pi}/G_{0}$, and the Hamiltonian saturates the BPS relation, which
is expressed by the sum of the F1 charge and the D0 tension:
\begin{eqnarray}
H&\equiv &\int^{\infty}_{-\infty}
dx\sqrt{G_{0}}\, (-{\hat T}^{0}_{\;0})\\
&=& \int^{\infty}_{-\infty}dx\sqrt{G_{0}}\, {\hat \Pi} +{\hat
{\cal T}}_{0}(1 + \hat E\theta_0). \label{ha00}
\end{eqnarray}
The energy density of D0 is localized near $x=0$ as
\begin{equation}\label{lide}
-{\hat T}^{0}_{\;0}={\hat \Pi}+ \frac{1}{\sqrt{G_{0}}}{\hat {\cal
T}}_{0}(1 + \hat E\theta_0) \frac{\xi/\pi}{x^{2}+\xi^{2}},
\end{equation}
where the width $\xi$ is
\begin{equation}
\xi=\frac{{\hat \Pi}}{{\hat E}} \frac{\sqrt{{\hat E}}R(1 + \hat
E\theta_0)}{{\hat {\cal T}}_{1}}.
\end{equation}

(i) The limit of ordinary EFT is achieved by making $\theta_{0}$
or equivalently $E_{0}$ vanish with fixed ${\hat E}$ and ${\hat
\Pi}$. (ii) If we take the limit of pure NCFT (${\hat
E}\rightarrow 0$), the open string metric becomes singular,
$G_{0}\rightarrow 0$ with keeping the closed
string variables finite, $G_{0}\theta_{0}={\hat
E}\theta_{0}=E_{0}/g_{0}= 1$. Then the pressure diverges, $-{\hat
T}^{11}\propto 1/\sqrt{G_{0}}\rightarrow\infty$, and the F1 charge
density vanishes, ${\hat \Pi}=-{\hat T}^{11}G_{0} \rightarrow 0$.
On the other hand, the tachyon configuration is smooth with finite
slope
\begin{equation}
{\hat \tau}(x)=\pm\frac{{\hat {\cal T}}_{1}}{2\sqrt{G_{0}}(-{\hat
T}^{11})}x,
\end{equation}
and its tension ${\hat {\cal T}}_{0}(1 + \hat E\theta_0)$
is also finite when ${\hat {\cal T}}_{1}$ is
finite
\begin{equation}
{\hat {\cal T}}_{0}(1 + \hat E\theta_0)
=\pi R{\hat {\cal T}}_{1}(1+{\hat E}\theta_{0})
=2\pi R {\hat {\cal T}}_{1}.
\end{equation}
This result is noteworthy because the codimension-one D-brane is
given by a thick smooth topological BPS object with finite tension
despite an infinite NC parameter, $\theta_{0}\rightarrow \infty$,
and singular open string metric, $G_{0}\rightarrow 0$. In terms of
closed string variables $g_{0}$ and $E_{0}$, this limit can be
understood as that of critical electric field $E_{0}=g_{0}$ for
nonvanishing $g_{0}$. Finiteness of the tension $2{\hat {\cal
T}}_{0}$ means vanishing
tension ${\cal T}_{0}$ in the EFT
due to Eq.~(\ref{toch}) and the relation (\ref{opc+}).
(iii) For finite $G_{0}$ or equivalently for finite ${\hat E}$,
the F1 charge density ${\hat \Pi}$ governs the width of kink
$\xi$: As ${\hat \Pi}$ goes to zero the localized piece of the
energy density (\ref{lide}) approaches a $\delta$-function for finite
tension, while large ${\hat \Pi}$ broadens the width.

\subsection{Topological kink for ${\hat \omega}^{2}<0$ (${\hat E}>G_{0}>0$)}
When ${\hat \omega}^{2}<0$, the potential $U_{1}({\hat \tau})$ in
Eq.~(\ref{potU}) becomes upside down leading to a hyperbolic solution.
Since the metric component $G_{0}$ in Eq.~(\ref{opc+}) is nonnegative,
we examine Eq.~(\ref{fde}) for the range ${\hat E}>G_{0}>0$.
For this range, the solution of the equation (\ref{fde}) is
\begin{equation}\label{hysi}
\frac{\hat\tau (x)}{R} = \pm \sqrt{\frac{{\hat {\cal T}}_1^2}{ (
\hat E^2 -G_{0}^{2})(-{\hat T}^{11})^{2}} + 1}\, \sinh (|{\hat
\omega}| x).
\end{equation}
The reason why we obtain a hyperbolic configuration is clearly
seen by looking at the action (\ref{ncfa01}) with a rescaling of $x$
\begin{eqnarray}
{\hat S}&=&-\int dt \sqrt{G_0}{\hat {\cal T}}_{1}(1 + \hat
E\theta_0) \int_{-\infty}^{\infty}dx\, {\hat V}({\hat \tau})
\sqrt{{\hat V}^{2}{\hat \tau}'^{2} -|{\hat \omega}^{2}|R^{2}}
\label{act5}\\
&=&-\int dt \sqrt{G_0}{\hat {\cal T}}_{1}(1 + \hat E\theta_0)
\int_{-\infty}^{\infty}d\eta\, {\hat V}\sqrt{{\hat V}({\hat
\tau})^{2}\left( \frac{d{\hat \tau}}{d\eta}\right)^{2} -1}
\nonumber \\
&\stackrel{{\hat \tau}/R=\sinh ({\hat T}/R)}{=}& -\int dt
\sqrt{G_0}{\hat {\cal T}}_{1}(1 + \hat E\theta_0)
\int_{-\infty}^{\infty}d\eta\,  V(\hat T) \sqrt{\left(
\frac{d{\hat T}}{d\eta}\right)^{2} -1} \, , \label{hrea}
\end{eqnarray}
where
\begin{equation}\label{eta}
\eta = |{\hat \omega}|Rx=\sqrt{\frac{\hat E^2-G_0^2}{G_0 (1 + \hat
E\theta_0)^2} }\, x .
\end{equation}
The rescaled action (\ref{hrea}) resembles formally that for a
time-dependent rolling tachyon
solution~\cite{Kim:2003ma,Mukhopadhyay:2002en} once we identify
$\eta$ as time and $\hat T$ as $T$ in EFT. Since $(d{\hat
T}/d\eta)^{2}-1$ occurs in the square root instead of $1-(d{\hat
T}/d\eta)^{2}$, hyperbolic sine solution (\ref{hysi}) is only
allowed.

Again the tension is computed from the action by substituting the
solution (\ref{hysi})
\begin{eqnarray}
\frac{{\hat S}}{-\int dt\sqrt{G_{0}}}&=& \frac{{\hat E}{\hat {\cal
T}}_{1}^{2}}{{\hat \Pi}G_{0}}
\int_{-\infty}^{\infty}dx \sqrt{G_{0}}\, {\hat V}^{2}\\
&=&2R{\hat {\cal T}}_{1}(1 + \hat E\theta_0)\, {\rm arctan}
\left(\frac{{\hat E}{\hat {\cal T}}_{1}}{{\hat \Pi} \sqrt{\hat E^2
-G_{0}^{2}}}\right), \label{eet}
\end{eqnarray}
which is less than $\pi R{\hat {\cal T}}_{1}(1 + \hat E\theta_0)$
and approaches this maximum value as $\hat E^2 \rightarrow
G_{0}^{2}$ or $\hat \Pi \rightarrow 0$. Note that $\sqrt{-X}$ from
Eq.~(\ref{X01}) or equivalently the square root in
Eq.~(\ref{hrea}) is always kept to be real and finite for making the kink
solution (\ref{hysi}) and the action (\ref{act5}) meaningful. The energy
density (\ref{t00}) is divided by a constant part from F1 fluid
and a localized piece from the ${\hat V}^{2}$  term
\begin{equation}\label{shne}
-{\hat T}^{0}_{\;0}={\hat \Pi}\frac{{\hat E}}{G_{0}} +\frac{{\hat
E}/G_{0}}{{\hat \Pi}} \frac{{\hat {\cal T}}_{1}^{2}}{
1+\left[\frac{{\hat {\cal T}}_{1}^{2}}{{\hat \Pi}^{2}\left[
1-\left(\frac{G_{0}}{{\hat E}}\right)^{2}\right]}+1\right]
\sinh^{2}\left[\sqrt{\frac{{\hat E}^{2}-G_{0}^{2}}{G_{0}(1+{\hat
E} \theta_{0})^{2}}}\,\frac{x}{R}\right] }.
\end{equation}
Despite the infinite slope of the tachyon profile $\hat\tau(x)$
(\ref{hysi}) at $x=\pm \infty$, the localized D$(p-1)$ part of
energy density (\ref{shne}) decreases exponentially to zero at the
asymptotic regions. The tension computed from the localized piece
of $- \hat T^{0}_{~0}$ (\ref{shne}) coincides exactly with the
value in Eq.~(\ref{eet}). Though its existence seems unconventional
due to the value of electric field ${\hat E}$ larger than the critical value,
the obtained kink has correspondence with a topological kink in
Ref.~\cite{Kim:2003in}.

Let us discuss various limits in what follows. (i) The limit of EFT is
smoothly taken by $\theta_{0}\rightarrow 0$ with fixed $G_{0}$,
which corresponds to $E_{0}\rightarrow 0$ with fixed
$g_{0}=G_{0}$. (ii) Pure NCFT limit is achieved in vanishing
interpolating field limit ${\hat E}\rightarrow 0$. Then ${\hat
E}>G_{0}>0$ condition leads to $G_{0}\rightarrow 0$. Since
consistency asks ${\hat E}^{2}/G_{0}\rightarrow 0$, ${\hat
\omega}^{2}=(G_{0}-{\hat E}^{2}/G_{0})/R^{2}\rightarrow 0^{-}$.
Therefore, it reduces to the same ${\hat E}\rightarrow 0$ limit of
the previous subsection 3.2, and the object is nothing but the thick
topological BPS kink in pure NCFT with $\theta_{0}\rightarrow
\infty$ and ${\hat E}\theta_{0}=1$. (iii) Finally we take thin
limit by taking ${\hat \Pi}\rightarrow 0$ for ${\hat E}>0$. Then
we have $-{\hat T}^{11}={\hat \Pi}/{\hat E} \rightarrow 0$ which
means the coefficient in front of hyperbolic sine function in
Eq.~(\ref{hysi}) is singular but the slope inside it remains finite. In
the expression of energy density (\ref{shne}), constant
contribution of the F1 vanishes and a sharply peaked localized
piece cannot become a $\delta$-function, while its tension recovers
${\hat \tau}_{0} =\pi R {\hat \tau}_{1}(1+{\hat E}\theta_{0})$.

We have obtained, from an unstable D1-brane, all possible static
codimension-one solitons identified as D0-branes. Let us finish this
section by summarizing the obtained kinks in a
table.
\begin{center}
\renewcommand{\arraystretch}{1.4}
\begin{tabular}{|c | c |c|} \hline
range of parameters & soliton species & type of solution \\ \hline
${\hat \omega}^{2}>0$ & array of kink-antikink & sinusoidal \\
${\hat \omega}^{2}=0$ & topological kink (BPS) & linear \\
${\hat \omega}^{2}<0$: ${\hat E}> G_{0}>0$ & topological kink with
${\hat \tau}(\pm\infty)=\pm\infty$
& hyperbolic sine
\\ \hline
\end{tabular}
\end{center}
\begin{center}{
Table 1: List of static solitons of codimension-one.}
\end{center}

\setcounter{equation}{0}
\section{Conclusion and Discussion}

In this paper we considered a real NC tachyon coupled to an NC
U(1) gauge field
describing NCFT version of the dynamics of an unstable D$p$-brane,
and proposed its action as DBI type which is
different from quadratic one already studied.
For slowly varying NC gauge field and NC tachyon, we showed
up to the leading NC parameter, the
equivalence between the proposed NC tachyon action and
the DBI type effective action of ordinary tachyon field.

For a flat unstable D1-brane with arbitrary diagonal component of
open string metric $G_{0}$, NC parameter $\theta_{0}$,
and interpolating electric field ${\hat E}$, we found all the static
kinks solutions, i.e., they are
array of kink-antikink and single topological kink.
The existence of kink solutions is
universal since they are supported irrespective of any ambiguity in assigning
star products between the NC fields, symmetrization procedure among
the star-producted terms, and
detailed shape of runaway NC tachyon potential.
For a specific tachyon potential, the kinks are given as exact solutions
of which functional forms coincide exactly with BCFT tachyon profiles.
Computing the tension of unit kink (or unit antikink), the obtained kinks
are identified with array of D0${\bar {\rm D}}$0 or single D0.

When $G_{0}^{2}={\hat E}^{2}$, there exists single topological kink
saturating BPS bound despite its nonzero thickness.
In the limit of singular open string metric $G_{0}\rightarrow 0$ or that of
divergent NC parameter $\theta_{0}\rightarrow\infty$ due to
$G_{0}\theta_{0}=1$, finiteness of the tension of NC kink requires
vanishing tension of kink in ordinary EFT.

The present study of NC kinks demonstrate clearly a relation for
D0 and their composites from an unstable D1 among NCFT, ordinary
EFT~\cite{Lambert:2003zr,Kim:2003in,Brax:2003rs,Kim:2003ma},
and BCFT~\cite{Sen:2003bc}. Therefore, further study on NC tachyon
is needed in terms of other languages
for off-shell string theory calculation, {\it e.g.}, string field
theories~\cite{Witten:2000nz}.

Though our discussion was restricted to the case of D1-brane, for
pure electric case with parallel ${\hat {\bf E}}$ and NC parameter
$\vec{\theta}$, an extension to D$p$-brane of arbitrary $p$ is
straightforward by choosing the transverse direction to
D$(p-1)$-brane as that of ${\hat {\bf E}}$. For the general
flat unstable D$p$-branes with various NC parameters
including spatial $\theta^{ij}$'s, the analysis becomes complicated.
It appears therefore that the D2 example might provide some useful
hints towards the solution of the more general problem.

As we made a smooth bridge from the kinks in ordinary EFT to those
in pure NCFT by employing the interpolating electric field ${\hat
{\bf E}}$, the same relation can be constructed for other tachyon solitons
like thick tachyon tubes~\cite{Kim:2003uc} or homogeneous time dependent
solutions represented by rolling tachyons~\cite{Mukhopadhyay:2002en}.
For spike (or BIon) configurations, thin NC solutions are
known~\cite{Hashimoto:2000mt} but any thick tachyon spike solutions are absent
both in EFT and NCFT up to now.

\section*{Acknowledgements}
The authors would like to thank J.L. Karczmarek, Chanju Kim, Chong Oh Lee,
Sangmin Lee, Y. Michishita, and A. Sen for valuable discussions. This work was
supported by funds from Sungkyunkwan University (R.B.) and is the
result of research activities (Astrophysical Research Center for
the Structure and Evolution of the Cosmos (ARCSEC)) supported by
Korea Science $\&$ Engineering Foundation (Y.K. and O.K.).

\appendix

\setcounter{equation}{0}
\renewcommand{\theequation}{A.\arabic{equation}}

\section{Equivalence between DBI action with tachyon and
NCDBI action with NC tachyon }

In this appendix we prove the equivalence between the DBI-type tachyon
effective action and its NC version given in Eq.~(\ref{ncfa}) up to
leading order in NC parameter $\theta$. The proof is similar
in spirit to that carried out in \cite{Seiberg:1999vs} for the non-tachyonic models.

DBI type tachyon effective action in the presence of the NS-NS
field on an unstable D$p$-brane is given in Eq.~(\ref{fa}).
For slowly varying
NC fields, $\hat F_{\mu\nu}$ and $\hat D_\mu\hat
T$, the NC counterpart of the action (\ref{fa})
is proposed in Eq.~(\ref{ncfa}).
Here we consider the general case including an interpolating field $\Phi_{\mu\nu}$,
analogous to the case without the NC tachyon (\ref{ncbi});
\begin{equation}\label{ncfa-A}
\hat S
= -\frac{1}{G_s(2\pi)^{\frac{p-1}{2}}}\int d^{p+1}x\, \hat
V(\hat T)\,\sqrt{-{\hat{X}}_{\Phi}},
\end{equation}
where
\begin{eqnarray}
{\hat{X}}_{\Phi}
= \det \left(G_{\mu\nu} +
\Phi_{\mu\nu} + \hat F_{\mu\nu} +\hat D_\mu \hat T \hat D_\nu \hat T \right).
\end{eqnarray}
The variables in the two actions (\ref{fa}) and
(\ref{ncfa-A}) are related by
\begin{eqnarray}
&&\frac{1}{G + \Phi} + \theta = \frac{1}{g + B}, \label{ocl1} \\
&& G_s = g_{{\rm s}} \sqrt{\frac{-\det (G + \Phi)}{-\det (g +
B)}}. \label{ocl2}
\end{eqnarray}
Using the SW map for the gauge
field $\hat A$~\cite{Seiberg:1999vs} and its extension to
NC tachyon field $\hat T$, the two actions (\ref{fa}) and
(\ref{ncfa-A}) will be shown to satisfy the following relation
\begin{equation}\label{rel}
 {\cal L} = \hat {\cal L} + \mbox{total}~~\mbox{derivative}.
\end{equation}
In order to prove the relation (\ref{rel}), we perform
a transformation of NC parameter,
$\theta \rightarrow \theta + \delta \theta$, for fixed $g_{{\rm
s}}$, $g_{\mu\nu}$ and $B_{\mu\nu}$, and see the small variation of
the action (\ref{ncfa-A}). The SW maps for the gauge
and real scalar fields are given
by~\cite{Seiberg:1999vs, Rivelles:2002ez,Banerjee:2004rs}
\begin{eqnarray}
\hat A_\mu &=& A_\mu - \frac{1}{4}\theta^{\kappa\lambda}\{
A_\kappa,~ \partial_\lambda A_\mu + F_{\lambda\mu}\} + {\cal
O}(\theta^2),
\label{SW-A} \\
\hat F_{\mu\nu} &=& F_{\mu\nu} + \frac{1}{4}
\theta^{\kappa\lambda}\left(2 \{F_{\mu\kappa},~ F_{\nu\lambda}\} -
\{ A_\kappa,~ D_\lambda F_{\mu\nu} + \partial_\lambda F_{\mu\nu}\}
\right) + {\cal O}(\theta^2), \label{SW-F}\\
\hat T &=& T - A_\alpha \theta^{\alpha\beta}\partial_\beta
T + {\cal O}(\theta^2), \label{SW-T} \\
\hat D_{\mu} \hat T &=& \partial_\mu T -
F_{\mu\alpha}\theta^{\alpha\beta}\partial_\beta T - A_\alpha
\theta^{\alpha\beta}\partial_\mu \partial_\beta T + {\cal
O}(\theta^2). \label{SW-DT}
\end{eqnarray}

Variation of Lagrange density $\hat {\cal L}$ in Eq.~(\ref{ncfa-A}) with
respect to $\theta$ is given by
\begin{equation}\label{varac}
\delta \hat {\cal L} =  \left[ -
\frac{\delta G_s}{G_s} + \frac{1}{\hat V} \delta \hat V
+ \frac{1}{2} \mbox{Tr} \left(
\frac{1}{\hat X}\left(\delta G + \delta \Phi + \delta \hat F +
\delta (\hat D \hat T \hat D \hat T)\right)\right)\right]\hat {\cal L},
\end{equation}
where we used matrix notation such as $(AB)_{\mu\nu} = A_{\mu\lambda}
B^{\lambda}_{~~\nu}$ and $\mbox{Tr} (AB) = A_{\mu\lambda}
B^{\lambda\mu}$. From Eqs.~(\ref{ocl1})--(\ref{ocl2}),
we have
\begin{eqnarray}
& & \delta G_s = \frac{1}{2} g_{{\rm s}} \sqrt{\frac{-\det (G +
\Phi)}{-\det ( g + B)}} \,\mbox{Tr}\left(\delta\theta (G +
\Phi)\right) = \frac{1}{2} G_s \mbox{Tr}\left(\delta\theta (G +
\Phi)\right) \label{var-Gs}, \\
& & \delta (G + \Phi) = ( G + \Phi) \delta\theta ( G + \Phi).
\label{var-GP}
\end{eqnarray}
The SW maps for the NC fields in Eqs.~(\ref{SW-A})--(\ref{SW-DT})
lead to
\begin{eqnarray}
\delta \hat F_{\mu\nu} &=& - \delta \theta^{\alpha\beta}(\hat F_{\mu\alpha}
\ast \hat F_{\beta\nu} ) - \frac{1}{2} \delta\theta^{\alpha\beta} \hat A_\alpha
\ast (\hat D_\beta \hat F_{\mu\nu} + \partial_\beta \hat F_{\mu\nu}),
\label{var-F}\\
\delta \hat T &=& - \frac{1}{2} \delta\theta^{\alpha\beta}(\hat A_\alpha \ast
\partial_\beta \hat T + \partial_\beta \hat T \ast \hat A_\alpha),
\label{var-T} \\
\delta\hat D_\mu \hat T &=& - \frac{1}{2} \delta \theta^{\alpha\beta}
\left( \hat F_{\mu\alpha}\ast \partial_\beta \hat T  + \partial_\beta \hat T \ast
\hat F_{\mu\alpha} + \hat A_\alpha \ast \partial_\mu\partial_\beta \hat T
+ \partial_\mu \partial_\beta \hat T \ast \hat A_\alpha \right),
\label{var-DT}
\end{eqnarray}
where we used the following property of star product
\begin{equation}
\delta \theta^{\mu\nu} \frac{\partial}{\partial\theta^{\mu\nu}}
(\hat f \ast \hat g)= \frac{1}{2} \delta^{\mu\nu} \partial_\mu \hat f
\ast \partial_\nu \hat g.
\end{equation}
Since proving the relation (\ref{rel}) up to the leading order in $\theta$
is of our interest, insertion of Eqs.~(\ref{SW-A})--(\ref{SW-DT})
into Eqs.~(\ref{var-F})--(\ref{var-DT}) results in
\begin{eqnarray}
\delta \hat F_{\mu\nu} &=& - (F\delta \theta F)_{\mu\nu} -
A_\alpha \delta\theta^{\alpha\beta}\partial_\beta F_{\mu\nu}
+ {\cal O}(\theta^2), \label{var-NC1} \\
\delta \hat T &=& - A_\alpha
\delta\theta^{\alpha\beta}\partial_\beta T
+ {\cal O}(\theta^2), \label{var-NC2} \\
\delta\hat D_\mu \hat T &=& - F_{\mu\alpha}\delta
\theta^{\alpha\beta} \partial_\beta T -
A_\alpha\delta\theta^{\alpha\beta}\partial_\mu\partial_\beta T
 + {\cal O}(\theta^2). \label{var-NC3}
\end{eqnarray}
Substituting Eqs.~(\ref{var-NC1})--(\ref{var-NC3}) into
Eq.~(\ref{varac}), we get
\begin{eqnarray}\label{var-L}
\delta \hat {\cal L} &=&  \left[
\frac{1}{\hat V}\frac{d \hat
V}{d\hat T} \delta \hat T- \frac{1}{2}
\mbox{Tr} \left(\delta (G + \Phi)\right) + \frac{1}{2} \mbox{Tr}
\left(\frac{1}{\hat X}( G + \Phi) \delta \theta ( G + \Phi)
\right)  \right. \nonumber \\
&& \hskip 0.4cm  \left. + \frac{1}{2} \mbox{Tr} \left(\frac{1}{\hat
X}\delta \hat F\right)+
\frac{1}{2} \left(\frac{1}{\hat X}\right)^{\nu\mu}
\left(\delta(\hat D_\mu \hat T) \hat D_\nu \hat T +\hat D_\mu
\hat T\delta( \hat D_\nu \hat T)\right)\right]\hat {\cal L}
\nonumber \\
&=&- \left[\frac{1}{\hat V} \frac{d\hat V}{d\hat T}
A_\alpha\delta \theta^{\alpha\beta} \partial_\beta T
+ \frac{1}{2}\mbox{Tr}(\hat F\delta\theta) -
\frac{1}{2} \mbox{Tr}\left( \frac{1}{\hat X} \hat F\delta\theta
 \hat D\hat T \hat D \hat T + \frac{1}{\hat X}
\hat D \hat T \hat D \hat T \delta\theta\hat F\right)
\right. \nonumber \\
&& \hskip 0.4cm + \frac{1}{2} \left(\frac{1}{\hat X}\right)^{\nu\mu} A_\alpha
\delta\theta^{\alpha\beta}\partial_\beta F_{\mu\nu}
+ \frac{1}{2} \left(\frac{1}{\hat X}\right)^{\nu\mu}
\left((F\delta\theta\partial T)_\mu\partial_\nu T +
\partial_\mu T (F\delta\theta\partial T)_\nu \right)
\nonumber \\
&& \hskip 0.4cm \left. + \frac{1}{2} \left(\frac{1}{\hat X}\right)^{\nu\mu}
\left(A_\alpha\delta\theta^{\alpha\beta}\partial_\beta
\partial_\mu T \partial_\nu T + A_\alpha\delta\theta^{\alpha\beta}
\partial_\beta \partial_\nu T \partial_\mu T\right)
\right] \hat {\cal L},
\end{eqnarray}
where in the second line we used the following relations
\begin{eqnarray}
\lefteqn{\mbox{Tr}\left(\frac{1}{\hat X}(G + \Phi) \delta\theta ( G +
\Phi) -\delta\theta ( G + \Phi)\right) +
\mbox{Tr}\left(\frac{1}{\hat X}\delta\hat F\right)} \nonumber \\
&& \,\, = - \mbox{Tr} \left(\frac{1}{\hat X}(\hat F
+ \hat D\hat T \hat D\hat T)
\delta\theta (\hat X - \hat F - \hat D\hat T \hat D\hat T)\right) -
\mbox{Tr}\left(\frac{1}{\hat X}\hat F \delta\theta \hat F\right)
\nonumber\\
&& \hspace{6mm}-
\left(\frac{1}{\hat X}\right)^{\nu\mu} A_\alpha \delta
\theta^{\alpha\beta}\partial_\beta F_{\mu\nu} \nonumber \\
&& \,\, = - \mbox{Tr} (\hat F \delta\theta)+ \mbox{Tr} \left(
\frac{1}{\hat X} \hat F \delta\theta \hat D\hat T \hat D\hat T
+ \frac{1}{\hat X} \hat D\hat T \hat D\hat T \delta\theta \hat F\right)
- \left(\frac{1}{\hat X}\right)^{\nu\mu} A_\alpha \delta
\theta^{\alpha\beta}\partial_\beta F_{\mu\nu}, \nonumber
\end{eqnarray}
and,
\[
\mbox{Tr} \left( \frac{1}{\hat X} \hat F \hat D\hat T \hat D\hat T
\delta\theta \hat D\hat T \hat D\hat T\right) = 0.
\]
Up to the leading order in $\theta$, we obtain
\begin{eqnarray}
\lefteqn{\mbox{Tr}\left( \frac{1}{\hat X} \hat F \delta\theta
\hat D\hat T \hat D\hat T+ \frac{1}{\hat X} \hat D\hat T \hat D\hat T
\delta\theta \hat F\right)} \nonumber\\
&&= \left(\frac{1}{\hat X}\right)^{\nu\mu}
\frac{\left(F\delta\theta \partial T\right)_\mu \partial_\nu T
+ \partial_\mu T \left(F\delta\theta \partial T\right)_\nu}{
1 + (\hat T /R)^2} + {\cal O}(\theta^2), \nonumber \\
\lefteqn{A_\alpha\delta\theta^{\alpha\beta}\partial_\beta \hat {\cal L}}
\nonumber\\
&&= \left[\left( \frac{1}{\hat V}\frac{d \hat V}{d\hat T}\right)
A_\alpha\delta\theta^{\alpha\beta} \partial_\beta \hat T
+ \frac{1}{2} \left(\frac{1}{\hat X}\right)^{\nu\mu}
A_\alpha\delta\theta^{\alpha\beta}\partial_\beta \hat F_{\mu\nu} \right.
\nonumber \\
&&\hskip 6mm \left.+ \frac{1}{2}\left(\frac{1}{\hat X}\right)^{\nu\mu}
\left( A_\alpha\delta\theta^{\alpha\beta}\partial_\beta
(\hat D_\mu \hat T) \hat D_\nu \hat T + \hat D_\mu \hat T
A_\alpha\delta\theta^{\alpha\beta}\partial_\beta (\hat D_\nu \hat T)
\right)\right] \hat {\cal L}.
\nonumber \\
&&
= \left[\left( \frac{1}{\hat V}\frac{d \hat V}{d\hat T}\right)
A_\alpha\delta\theta^{\alpha\beta} \partial_\beta  T
+ \frac{1}{2} \left(\frac{1}{\hat X}\right)^{\nu\mu}
A_\alpha\delta\theta^{\alpha\beta}\partial_\beta  F_{\mu\nu} \right.
\nonumber \\
&&\hskip 4mm \left.+ \frac{1}{2}\left(\frac{1}{\hat X}\right)^{\nu\mu}
\left( A_\alpha\delta\theta^{\alpha\beta}\partial_\beta
( \partial_\mu  T) \partial_\nu  T + \partial_\mu  T
A_\alpha\delta\theta^{\alpha\beta}\partial_\beta (\partial_\nu  T)
\right)\right] \hat {\cal L}
+ {\cal O}(\theta^2).
\label{der-L}
\end{eqnarray}
Then, with the help of Eqs.~(\ref{var-L})--(\ref{der-L}), the variation of
Lagrange density (\ref{varac}) is expressed by total derivative terms
for leading order in $\theta$;
\begin{eqnarray}
\delta \hat {\cal L} &=& -\frac{1}{2} \hat {\cal L}
F_{\alpha\beta} \delta\theta^{\beta\alpha} - A_\alpha\delta\theta^{\alpha\beta}
\partial_\beta\hat {\cal L} + {\cal O}(\theta^2)
\nonumber \\
&=& -\partial_\beta(A_\alpha \delta\theta^{\alpha\beta} \hat {\cal L})
+ {\cal O} (\theta^2).
\label{var-L01}
\end{eqnarray}
This completes the proof of the relation (\ref{rel}) up to leading order
in $\theta$.

\section{Derivation of NC Equations of Motion}

\subsection{NC tachyon equation}
Let us consider variation of the NC action (\ref{ncfa}) when a small
variation of the NC tachyon field $\hat T$ is taken;
\begin{equation}\label{delS}
\hat S[\hat T + \delta\hat T] - \hat S[\hat T] = \int
d^{p+1}x\, \frac{\delta\hat S[\hat T]}{\delta\hat T(x)} \delta
\hat T (x)
= \delta_{\hat T} \hat S_1 [\hat T] +
\delta_{\hat T} \hat S_2[\hat T],
\end{equation}
where
\begin{eqnarray}
\delta_{\hat  T} \hat S_1 [\hat T] &=& - \hat {\cal T}_p \int d^{p+1}x\,
\delta_{\hat T} \hat V (\hat T) \sqrt{- \hat X},
\label{delS1} \\
\delta_{\hat T} \hat S_2[\hat T] &=& - \hat {\cal T}_p \int d^{p+1}x\,
\hat V(\hat T)\, \delta_{\hat T} \sqrt{- \hat X}.
\label{delS2}
\end{eqnarray}
From the definition of the NC tachyon potential (\ref{ncpot}) we
obtain
\begin{eqnarray}\label{delV}
\delta_{\hat T} \hat V (\hat T) &=& \hat V(\hat  T +
\delta\hat T) - \hat V(\hat  T) \nonumber \\
&=& -\frac{1}{2} \left(\frac{\delta \hat T}{R} \ast
\frac{\hat T}{R} +\frac{\hat T}{R} \ast
\frac{\delta\hat T}{R}\right) \nonumber \\
&& + \frac{5}{24} \left(\frac{\delta\hat T}{R} \ast
\frac{\hat T}{R} \ast\frac{\hat T}{R}\ast \frac{\hat T}{R}
+\frac{\hat T}{R} \ast \frac{\delta\hat T}{R}
\ast\frac{\hat T}{R}\ast
\frac{\hat T}{R}\right. \nonumber \\
&&\left.\qquad +\frac{\hat T}{R} \ast \frac{\hat T}{R}
\ast\frac{\delta\hat T}{R}\ast \frac{\hat T}{R} +\frac{
\hat T}{R} \ast \frac{\hat T}{R} \ast\frac{\hat T}{R}\ast
\frac{\delta\hat T}{R}\right)  + \cdots
\nonumber \\
&=& - \left[ \frac{\sinh (\hat T/R) \delta \hat T}{R
\cosh^2 (\hat T/R)}\right]_\ast,
\end{eqnarray}
Substituting Eq.~(\ref{delV}) into Eq.~(\ref{delS1}) and
making use of the cyclic property of star
product, we rewrite Eq.~(\ref{delS1}) as
\begin{equation}\label{delS1-1}
\delta_{\hat T}\hat S_1 [\hat T] = {\cal T}_p  \int
d^{p+1} x\, \left[ \frac{\sinh (\hat T/R) }{
R \cosh^2 (\hat T/R)}\right]_\ast \sqrt{ - \hat X}\,  \delta \hat T.
\end{equation}

Inserting the formula of cofactor (\ref{dsX}) into
Eq.~(\ref{delS2}), we arrive at
\begin{eqnarray}\label{delS2-1}
\delta_{\hat T} \hat S_2[\hat T] &=& \frac{\hat {\cal T}_p}{2}
\int d^{p +1}x \, \frac{\hat V (\hat T)}{\sqrt{- \hat X}}
\hat C^{\mu\nu}\, \delta_{\hat T} \hat X_{\mu\nu},
\nonumber \\
 &=& \hat {\cal T}_p
\int d^{p +1}x \, \frac{\hat V (\hat T)}{\sqrt{- \hat X}}\,
\hat C^{\mu\nu}_{{\rm S}} \, \hat D_\mu\delta\hat T \hat D_\nu\hat T
\nonumber \\
&=& -\hat {\cal T}_p \int d^{p +1}x \,\hat D_\mu \left(
\frac{\hat V (\hat T)}{\sqrt{- \hat X}}\,
\hat C^{\mu\nu}_{{\rm S}} \hat D_\nu\hat T\right)\, \delta \hat T.
\end{eqnarray}
In the last step of Eq.~(\ref{delS2-1})
we have used the following property of the star product
\begin{equation}
\int \hat f \ast \hat D_\mu \hat g = \int \partial_\mu (\hat f
\ast \hat g) - \int (\hat D_\mu \hat f) \ast \hat g = - \int (\hat
D_\mu \hat f)  \hat g.
\end{equation}
Plugging Eqs.~(\ref{delS1-1})--(\ref{delS2-1}) into the variation of
the action (\ref{delS}),
we find the equation of motion for the NC tachyon field (\ref{ncte}).

\subsection{NC U(1) gauge field equation}

Similar to the previous section, we calculate the variation of NC action
under a small variation of the NC gauge field $\hat A_\mu$
\begin{eqnarray} \label{delSA}
\delta_{\hat A} \hat S[\hat A] &=& \hat S[\hat A + \delta \hat A]
- \hat S[\hat A] = \int d^{p+1} x\, \frac{\delta \hat S[\hat A]}{
\delta \hat A_\mu (x)} \delta \hat A_\mu (x) \nonumber \\
&=& -  \hat {\cal T}_p \int d^{p+1} x\,  \hat V \,
\delta_{\hat A} \sqrt{- \hat X} \\
&=& \frac{{\cal T}_p}{2} \int d^{p+1} x\, \frac{\hat V(\hat T)}{
\sqrt{-\hat X}} \, \hat C^{\mu\nu} \delta_{\hat A} \hat X_{\mu\nu}
\nonumber \\
&=&\delta_{\hat A} \hat S_1[\hat A] + \delta_{\hat A} \hat
S_2[\hat A],
\label{delSA01}
\end{eqnarray}
where
\begin{eqnarray}
\delta_{\hat A} \hat S_1[\hat A] &=& \frac{{\cal T}_p }{2} \int d^{p+1}
x\, \frac{\hat V(\hat T)}{
\sqrt{-\hat X}} \, \hat C^{\mu\nu}_{{\rm A}} \delta_{\hat A} \hat F_{\mu\nu},
\label{delSA1} \\
\delta_{\hat A} \hat S_2[\hat A] &=& {\cal T}_p \int d^{p+1}
x\,\frac{\hat V(\hat T)}{
\sqrt{-\hat X}} \, \hat C^{\mu\nu}_{{\rm S}}\,
\delta_{\hat A}\left(\hat D_\mu \hat T\right) \hat D_\nu \hat T.
\label{delSA2}
\end{eqnarray}
From the definition of NC gauge field strength (\ref{ncF}) we obtain
\begin{eqnarray} \label{delSA1-1}
\delta_{\hat A} \hat S_1[\hat A] &=& {\cal T}_p  \int d^{p+1} x\,
\frac{\hat V(\hat T)}{\sqrt{-\hat X}} \,
\hat C^{\mu\nu}_{{\rm A}} \hat D_\mu \delta \hat A_\nu
\nonumber \\
&=& - {\cal T}_p  \int d^{p+1} x\,
\hat D_\mu\left( \frac{\hat V(\hat T)}{\sqrt{-\hat X}} \,
\hat C^{\mu\nu}_{{\rm A}}\right) \delta \hat A_\nu,
\end{eqnarray}
and the definition of the covariant derivative (\ref{code}) leads to
\begin{eqnarray} \label{delSA2-1}
\delta_{\hat A} \hat S_2[\hat A] &=&
-i \hat {\cal T}_p \int d^{p +1}x\, \frac{\hat V(\hat T)}{
\sqrt{-\hat X}} \, \hat C^{\mu\nu}_{{\rm S}}\hat D_\nu \hat T \ast
\left( \delta\hat A_\mu \ast \hat T - \hat T \ast \delta \hat A_\mu
\right) \nonumber \\
&=& - i \hat {\cal T}_p \int d^{p +1}x \,
\left[ \hat T\, , \, \frac{\hat V(\hat T)}{
\sqrt{-\hat X}} \, \hat C^{\mu\nu}_{{\rm S}}\hat D_\nu \hat T \right]_\ast
\delta \hat A_\mu,
\end{eqnarray}
where we used
\begin{equation}
\delta_{\hat A} (\hat D_\mu \hat T) = -i (\delta \hat A_\mu
\ast \hat T - \hat T \ast \delta \hat A_\mu).
\end{equation}
Substituting Eqs.~(\ref{delSA1-1})--(\ref{delSA2-1}) into Eq.~(\ref{delSA01})
provides
the gauge field equation (\ref{ncge}).

\end{document}